\documentclass[%
superscriptaddress,
preprint,
showpacs,preprintnumbers,
showkeys
nobibnotes,
amsmath,amssymb,
prstab,
tightenlines,
showkeys,
floatfix,
]{revtex4-1}
\usepackage{graphicx}
\usepackage{revsymb4-1}
\usepackage{bm}
\usepackage{braket}
\usepackage{mathptmx}
\usepackage[colorlinks, linkcolor=blue, anchorcolor=blue, citecolor=blue]{hyperref}
\usepackage{hyperref}

\begin{document}

\title{Exploring quantum correlations in a hybrid optomechanical system}


\author{Smail Bougouffa \thanks{corresponding author}}
\email{sbougouffa@hotmail.com and sbougouffa@imamu.edu.sa}

 \author{Mohannad Al-Hmoud} 
\email{mmalhmoud@imamu.edu.sa}
\affiliation{ Department of Physics, College of Science, Imam Mohammad ibn Saud Islamic University (IMSIU), P.O. Box 90950, Riyadh 11623, Saudi Arabia }

\author{ Jabir Wali Hakami} 
\email{j.hakami@jazanu.edu.sa}
\affiliation{Department of Physics, College of Science, Jazan University, Jazan, Saudi Arabia}

\date{\today}

\begin{abstract}
 In quantum simulations and experiments on optomechanical cavities, coherence control is a challenging issue. We propose a scheme of two coupled optomechanical cavities to enhance the intracavity entanglement. Photon hopping is employed to establish couplings between optical modes, while phonon tunneling is utilized to establish couplings between mechanical resonators. Both cavities are driven by classical light. 
We explore the influences of coupling strengths of the quantum correlations generated inside each cavity using two types of quantum measures: logarithmic negativity and quantum steering. This analysis will reveal the significance of these quantum metrics as well as their various aspects in the Doppler regime. We also investigate stability conditions based on coupling strengths. Therefore, it is possible to quantify the degree of intracavity entanglement.  The generated entanglement can be enhanced by choosing the appropriate photon and phonon hopping strengths. 
A set of parameters based on the currently available experimental data was used in the calculations.



\end{abstract}

\maketitle

\section{Introduction}\label{sec1}

Hybrid optomechanical systems are utilized as broad schemes to describe the quantum characterizations of macroscopic mechanical resonators\cite{Aspelmeyer2008, Blencowe2004, Genes2009, Aspelmeyer2010, Clerk2014}.   They are utilized to generate quantum light-mechanical states and to carry out quantum information processing \cite{Liu2013, Yan2015, Yan2019} as well as to achieve precise measurements, and control \cite{LIGO2007, purdy2013observation}.

The transfer of quantum correlations from an entangled light source to initially separable or coupled cavity optomechanics, on the other hand, is well established and demonstrated in a variety of systems  \cite{Lee2005, Bougouffa2012, Bougouffa2013b, Sete:14,sete2015high, Yan2015}.

Due to the advancements in computation
 \cite{nielson2000quantum}, secure communication \cite{acin2007device}, and metrology \cite{giovannetti2011advances},  quantum correlations have indeed been extensively studied in recent years. 
Quantum correlations can appear in a variety of processes for mixed states of composite quantum systems \cite{modi2012classical}. 
However, while entanglement  \cite{horodecki2009quantum} is one of the most well-studied of these phenomena, quantum steering \cite{schrodinger1935discussion} is a transitional type of quantum correlation that has only recently stimulated the involvement of the quantum information community \cite{skrzypczyk2014quantifying}, introducing new possibilities for theoretical investigations and experimental applications.

In mesoscopic systems, quantum entanglement has been extensively studied \cite{Zhou2011,nunnenkamp2011single,Purdy2013a,Bai2016,Bougouffa2016,Liang2019,Ge2013a,Ge2015,Ge2015a,Si2017,Asiri2018}. Hybrid optomechanical systems have been presented as promising options for transferring entanglement from entangled light to mechanical resonators \cite{PhysRevA.85.043824,ElQars2017,Kronwald,Yousif2014}. As a consequence, the generated entangled states are known as “Schr\"{o}dinger cat" states, in which a "microscopic" degree of freedom (the optical cavity mode) is entangled with a "macroscopic" (or mesoscopic) degree of freedom (the mechanical resonator).

Since linear optomechanical cavities investigate the interaction of photons and phonons via radiation pressure, they are well suited to exploring experimentally and theoretically, various important quantum concepts, such as entanglement \cite{Yan2019, liao2022dissipation, lin2021enhancement}, ground state cooling \cite{Aspelmeyer2010}, quantum synchronization \cite{Ameri2019, Qiao2018}, quantum channel discrimination \cite{marchese2021optomechanical}, and others. Different connected optomechanical systems, on the other hand, have been proposed for these reasons \cite{Li2017, Paule2018}. Existing research in this area is mostly focused on alternative schemes of linking some optomechanical cavities. For quantum information protocols utilizing continuous variables systems, generated entanglement \cite{Yan2019} provides the conventional implementation.  

Moreover, the creation of several quantum states in a micromechanical resonator has already been studied. Squeezed and entangled states are the most significant examples. Squeezed states of mechanical resonators \cite{Jaehne2009} have the potential to go above the usual quantum limit for position and force detection \cite{jurcevic2021demonstration}, and they can be created in a variety of ways, including Optomechanical cavities \cite{Asjad2016}. Entanglement, on the other hand, is a distinguishing feature of quantum theory since it is responsible for observable correlations that cannot be explained by local realistic theories. Thus, there has been a growing interest in determining the conditions under which macroscopic objects can get entangled. However, there has been an increasing interest in figuring out what causes macroscopic objects to become entangled. One of the most promising candidates for this is the optomechanical cavity. 

The generated entanglement is quantified in virtually proposed optomechanical systems by the logarithmic negativity, which is measurable and quantifiable, and an upper bound to the distillable entanglement. It is also a criterion for separability derived from the PPT criterion \cite{Plenio}. To probe certain useful quantum features, it's interesting to look at the efficacy of alternative entanglement metrics in hybrid systems. It's also fascinating to investigate the fundamental purpose of other proposed metrics in physical systems.

The type of interactions between distinct modes in the optomechanical cavity has an impact on the generated entanglement in general.
We present a scheme of hybrid optomechanical systems that will allow us to investigate the origin of quantum correlations. Also, what form of interaction in the system can promote intracavity mode entanglement? Furthermore, we will be able to employ and analyze two alternative entanglement metrics using this suggested scheme, namely, logarithmic negativity and quantum steering between intracavity modes.

The paper is structured as follows. We present the system of two connected optomechanical cavities driven by classical light in Section  \ref{sec2} and establish the Hamiltonian for the system. The collective quantum dynamics are presented in the rotating wave approximation in section  \ref{sec3}. We derive the coupled stochastic differential equations for quadrature fluctuations using the linearization technique, which can be solved in the steady-state approximation. The correlation matrix is obtained in Section \ref{sec4}, and the stability requirements are explored. To quantify the degree of quantum correlations between optical and mechanical modes, two quantum measures, namely logarithmic negativity and quantum steering, are described in Section \ref{sec5}. We also go over the numerical results of various quantum measures. Our conclusions are given in Section \ref{sec6}.

\section{The system}\label{sec2}
\begin{figure}[ht]
\hspace*{2cm}\textbf{(a)} \hspace*{6.cm}\textbf{(b)}\\
\includegraphics[width=0.45\linewidth,height=0.40\linewidth]{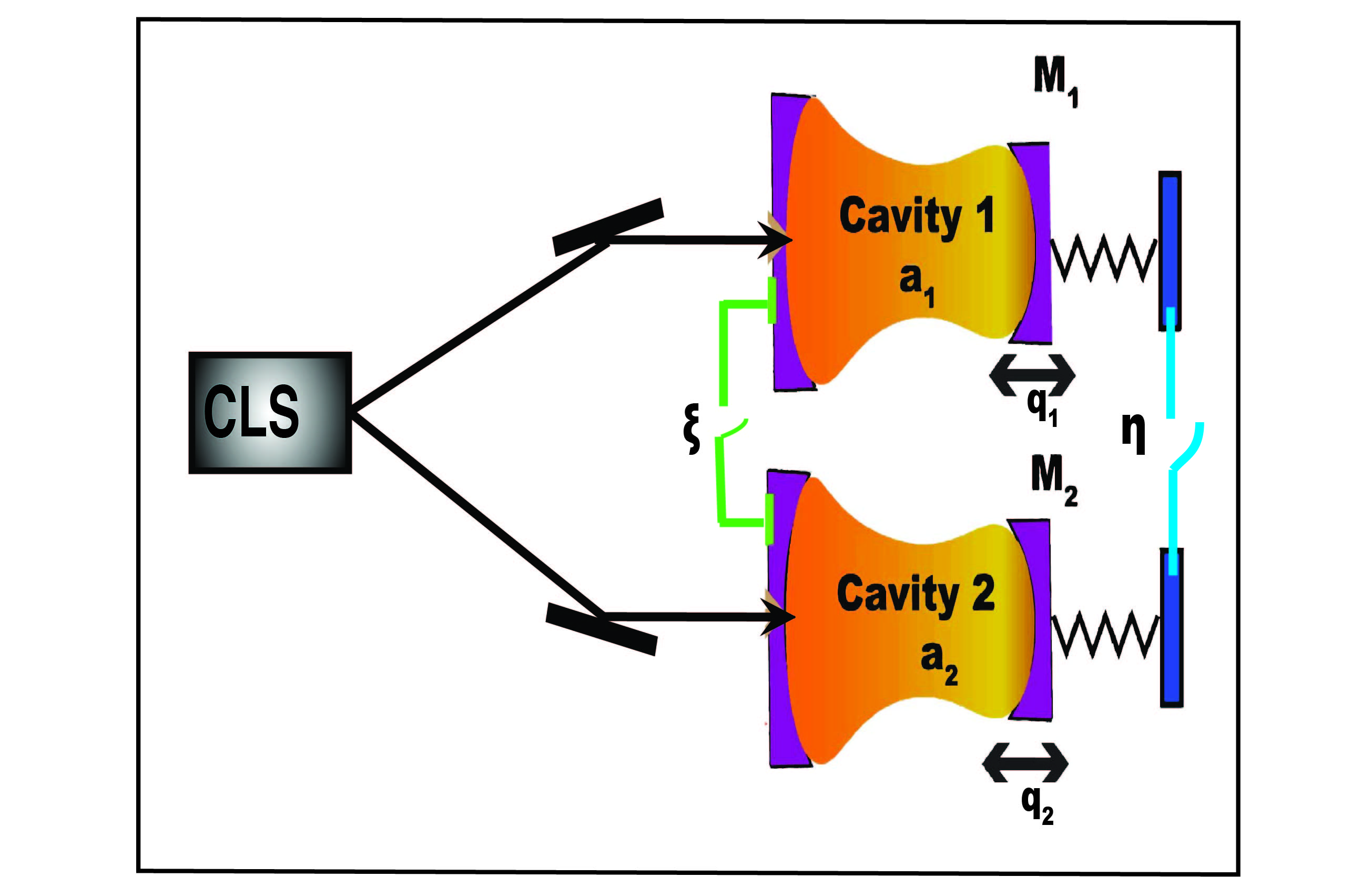}
\includegraphics[width=0.45\linewidth,height=0.45\linewidth]{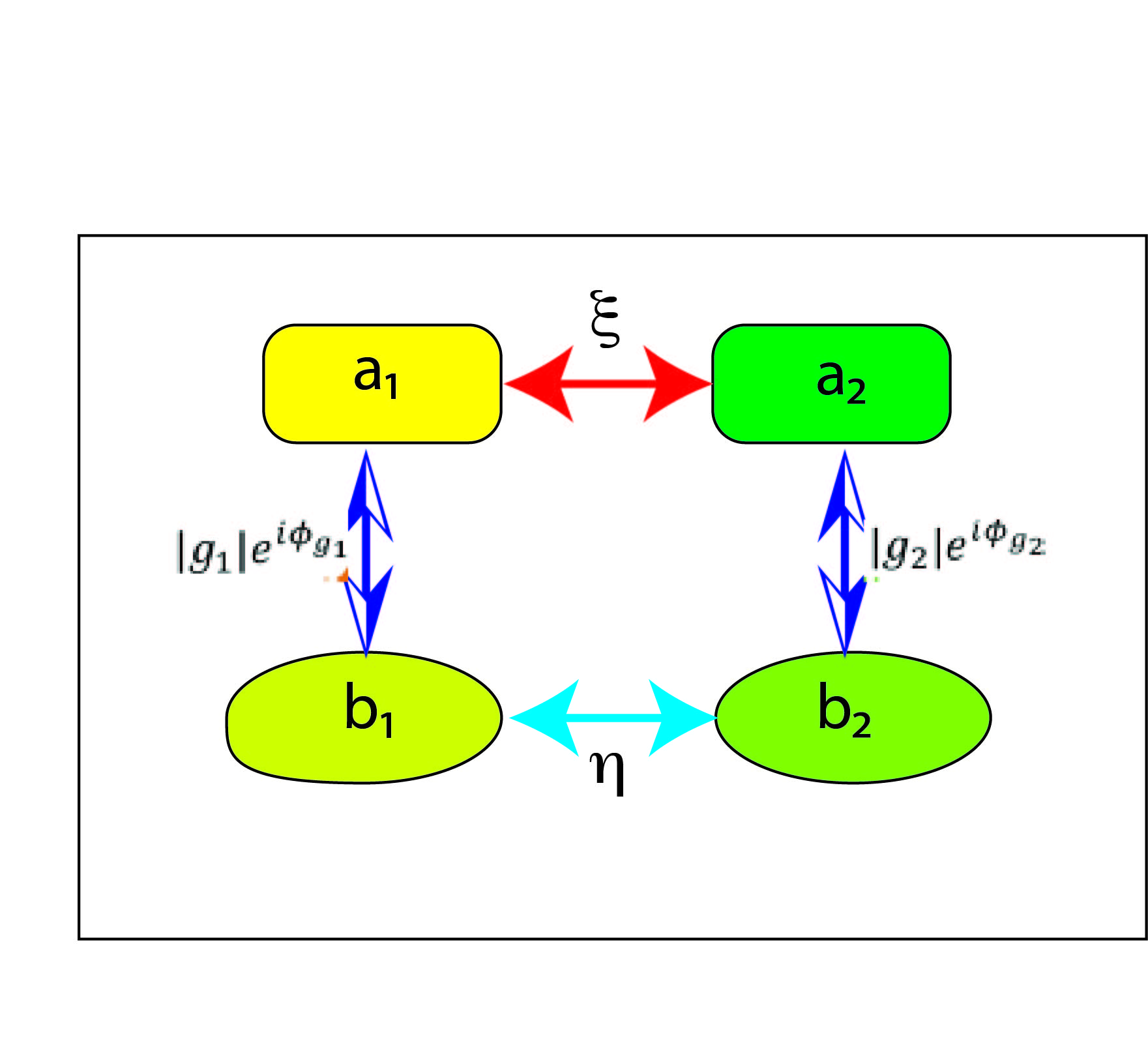}
\caption{(Color online)(a) Schematic diagram of the system to synchronize two phonon modes in a pair of single-mode optomechanical cavities 1 and 2. Both cavities are driven by coherent light source (CLS). Both photons and phonons can tunnel between each of them at rates $\xi$ and $\eta$, respectively.  (b) The different coupling strength and
phases are introduced between two modes of four.}\label{Fig1}
\end{figure}

Let us consider a system composed of two linked optomechanical cavities in this paper. An optical fiber ($xi$) establishes the first coupling between the fixed mirrors, i.e. a photon hopping process. The output field of a coherent light source is exposed to the fixed sides (CLS). A phonon tunneling mechanism ($eta$) \cite{Ludwig2013}  is used to introduce the second coupling between the moving mirrors, as shown in Fig.\ref{Fig1}. 
 
Through the rotating wave approximation in an acceptable observation context, the Hamiltonian describing the scheme's unitary dynamics reads ($\hbar=1$)
\begin{eqnarray}\label{eq1}
 \hat{H}&=&\sum_{j=1,2}\Big[\Delta_{0j} a_{j}^\dagger a_{j}
         + \omega_{mj}b_j^\dagger b_j
         - g_{j} a_{j}^\dagger a_{j} (b_j^\dagger+ b_j)\nonumber \\&{}&
         +i(E_j a_{j}^\dagger-E_j^{*} a_{j} )\Big]-\eta (b_1^\dagger b_2+b_2^\dagger b_1) -\xi \big( a_{1}^\dagger a_{2}+a_{2}^\dagger a_{1}\big).
\end{eqnarray}
The cavity mode $(a_j )$ is transformed into the frame rotating at the laser frequency $(\Delta_{0j}=\omega_j-\omega_L)$ and driven at the rate $(E_j)$. The mechanical mode $(b_j)$ is characterized by a frequency $\omega_{mj}$. In the most general case, both photons and phonons can tunnel between each of them at rates $(\xi)$ and $(\eta)$, respectively. These cavities consist of a mechanical mode and a laser-driven optical mode that interact via the radiation pressure coupling at a rate $(g_j)$. The quantum dynamics of the suggested system can be presented in terms of the quantum Langevin equations.

\section{Collective quantum dynamics}\label{sec3}

The fluctuation-dissipation processes influencing both the cavity and the mechanical modes determine the analysis of the proposed system's collective quantum dynamics. The following set of nonlinear quantum Langevin equations (QLEs) can be obtained using the Hamiltonian (\ref{eq1}) and dissipation processes, stated in the interaction picture for $\sum_{j=1,2}\hbar\omega_{L j} a_{j}^{\dag}a_{j}$, 
\begin{eqnarray}\label{eq2}
    \dot{b_{1}}&=&-\beta_1 b_{1} + ig_{1} {a_{1}}^{\dag}a_{1}+i\eta b_2+\sqrt{2\gamma_{1}}b^{in}_{1}, \nonumber\\
        \dot{b_{2}}&=&-\beta_2 b_{2} + ig_{2} {a_{2}}^{\dag}a_{2}+i\eta b_1+\sqrt{2\gamma_{2}}b^{in}_{2}, \nonumber\\
    \dot{a_{1}}&=&-\alpha_1 a_{1} +i g_{1} a_{1} (b_1^\dagger+ b_1) +i \xi a_{2}+E_1 +\sqrt{2\kappa_{1}} a^{in}_{1},\nonumber\\
     \dot{a_{2}}&=&-\alpha_2 a_{2} +i g_{2} a_{2} (b_2^\dagger+ b_2) +i \xi a_{1}+E_2 +\sqrt{2\kappa_{2}} a^{in}_{2},\nonumber\\
\end{eqnarray}
where $\beta_{j}= \gamma_j+i\omega_{mj}, \alpha_j= \kappa_j+i\Delta_{0j}$ for ($j=1,2$). $ \kappa_j$ and $\gamma_{j}$  are the damping rates for the cavity and mechanical mode ($mj$), respectively. 
With $b_j^{in}$ being the thermal noise on the mechanical mode ($mj$) with zero mean value and  ($a_j^{in}$) being the input quantum vacuum noise from the cavity (j) with zero mean value.

In the Markovian approximation, the Brownian noise $b_j^{in}$ is characterized by the correlation functions \cite{giovannetti2001phase,Rehaily2017,Gardiner,Benguria,Zhang}
\begin{eqnarray}\label{eq5}
\big\langle b_j^{in}(t) {b_j^{in}}^{\dagger}(t')\big\rangle &=&\ (\bar{n_{j}}+1)\delta(t-t'),\nonumber\\
\big\langle {b_j^{in}}^{\dagger}(t) b_j^{in}(t')\big\rangle &=&\ \bar{n_{j}}\delta(t-t')
\end{eqnarray}
where $\bar{n_{j}} = (e^{\hbar \omega_{mj}/k_{B} T} - 1)^{-1} $ is the mean number of thermal phonons at the frequency of the mechanical resonator $j$, $k_b$ is the Boltzmann constant, and T is the temperature of the environment. 

The input quantum optical noise $a^{in}_{j}$ with zero mean value, satisfies the Markovian correlation functions \cite{Gao,Parkins1990}:
\begin{eqnarray}\label{eq3}
\langle a_j^{in}(t) {a_{j}^{in}}^{\dagger}(t')\rangle &=&(N_{j}(\omega_j)+1)\delta(t-t'),\nonumber\\
\langle {a_j^{in}}^{\dagger}(t) a_{j}^{in}(t')\rangle &=&N_{j}(\omega_j)\delta(t-t'),
\end{eqnarray}
where $N(\omega_j)=[\exp(\hbar \omega_j/k_BT)-1]^{-1}$ is the equilibrium mean thermal photon number of the optical field, which can be assumed $N(\omega_j)\approx 0$.

Equations (2), in conjunction with the correlation functions (3) and (4), characterize the dynamics of the system in question completely. Because of the interaction between the cavity modes and the two mechanical mirrors, these coupled equations have an important property: essential nonlinearity. Nonlinearity has an important role in achieving self-sustaining vibrations for phonon modes and their impact on transmitted photons.

The modes $a_1$ and $a_2$ are directly coupled to one other with the strength $xi$ and are also indirectly related to each other via the coupling to the corresponding vibrating mode with the strengths $g_1$ and $g_2$, respectively, as shown in the system (\ref{eq2}). As seen in figure 1(b), this coupling configuration creates a closed loop, and the dynamics of the system can then exhibit phase-dependent effects \cite{Sun2017}.

To get strong optomechanical coupling for cooling both the mechanical oscillators, we consider an intense laser pump, leading to large amplitude of the cavity field $| \langle a_j\rangle | \gg 1$. This permits us to linearize the system dynamics around the semiclassical averages \cite{Li2016, Braunstein2012, Li2015,Genes2009}  by decomposing each operator as $O = \langle O\rangle + \delta O$ $(O = a_j, b_j)$ with the factorization conjecture $\langle b_j a_j\rangle= \langle b_j\rangle\langle a_j\rangle$, and neglecting small second-order fluctuation terms. Therefore, the Eqs. (\ref{eq2}) can be spent into two parts of equations: one is for averages $O^{ss} = \langle O\rangle$ and the other for zero-mean quantum fluctuations $\delta O$. 

The steady-state mean values can be read as
\begin{eqnarray}\label{eq6}
b^s_{j}&=&\frac{ig_{j} \mid a^s_{j}\mid^{2}(\gamma_{3-j}+i\omega_{mj})-\eta g_{3-j} \mid a^s_{3-j}\mid^{2} }{(\gamma_j+i\omega_{mj})(\gamma_{3-j}+i\omega_{m{3-}j})+\eta^2}; \quad j=1,2,\nonumber\\
a^s_{j}&=&\frac{\alpha'_{3-j}E_j+i\xi E_{3-j}}{\alpha'_j\alpha'_{3-j}+\xi^2}; \quad j=1,2,
\end{eqnarray}
where $\alpha'_j=\kappa_j+i\Delta_j, (j=1,2).$
The last two equations of (\ref{eq6}) are indeed nonlinear equations providing the steady intracavity field amplitude $a^s_{j}$, as the effective cavity-laser detuning $\Delta_{j}$, including radiation pressure effects, which is given by $\Delta_j=\Delta_{0j}-2g_j\Re(b_j^s)$. The parameter regime appropriate for generating quantum correlation is that with a very large input power $P$, i.e., as $|a^s_{j}|\gg 1$. The nonlinearity also demonstrates that the stable intracavity field amplitude can reveal variable behavior for a given parameter range.

The  linearized QLEs for the quadrature fluctuations are given by
\begin{eqnarray}\label{eq8}
    \dot{\delta b_{1}}&=&-\beta_{1}\delta b_{1} + ig_{1}a_1^s  (\delta a_{1}^{\dag}+\delta  a_{1})+i\eta \delta b_2+\sqrt{2\gamma_{1}}b^{in}_{1}, \nonumber\\
    \dot{\delta b_{2}}&=&-\beta_{2}\delta b_{2} +i g_{2} a_2^s( \delta a_{2}^{\dag}+\delta  a_{2})+i\eta \delta b_1+\sqrt{2\gamma_{2}}b^{in}_{2}, \nonumber\\    
    \dot{\delta a_{1}}&=&-\alpha'_{1}\delta a_{1} +i g_{1} a_{1}^s (\delta b_1^\dagger+ \delta b_1) +i \xi \delta a_{2}+\sqrt{2\kappa_{1}} a^{in}_{1},\nonumber\\
     \dot{\delta a_{2}}&=&-\alpha'_{2}\delta a_{2} +i g_{2} a_{2}^s (\delta b_2^\dagger+ \delta b_2) +i \xi \delta a_{1}+\sqrt{2\kappa_{2}} a^{in}_{2},\nonumber\\
\end{eqnarray}
where $a_{j}^s$ is assumed to be real.

Taking the transformations of optical and mechanical mode operators
\begin{eqnarray}\label{eq9}
   \delta X_{j} = \frac{\delta a_{j} +\delta a_{j}^\dag}{\sqrt{2}},\quad \delta Y_{j} = \frac{\delta a_{j} -\delta a_{j}^\dag}{i \sqrt{2}},\\
   \delta q_{j} = \frac{\delta b_{j} +\delta b_{j}^\dag}{\sqrt{2}},\quad \delta p_{j} = \frac{\delta b_{j} -\delta b_{j}^\dag}{i \sqrt{2}},
\end{eqnarray}

and the analogous hermitian input quadrature noises
\begin{eqnarray}\label{eq10}
    X_{j}^{in} = \frac{a_{j}^{in} + {a_j^{in}}^{\dag}}{\sqrt{2}}, \quad  Y_{j}^{in}= \frac{ a_{j}^{in} - {a_j^{in}}^{\dag}}{i \sqrt{2}},
\end{eqnarray}
\begin{eqnarray}\label{eq10a}
    Q_{j}^{in} =\frac{ b_{j}^{in} + {b_j^{in}}^{\dag}}{\sqrt{2}}, \quad   P_{j}^{in} =  \frac{ b_{j}^{in} - {b_j^{in}}^{\dag}}{i \sqrt{2}},
\end{eqnarray}
Then, Eq. (\ref{eq8}) can be written as
\begin{equation}
\label{eq11}
\dot{\boldsymbol{\mu}}(t)=A\boldsymbol{\mu}(t)+\mathbf{ C}(t),
\end{equation}
with the vector of quadrature fluctuations
\begin{equation}\label{eq12}
\boldsymbol{\mu}(t)=\Big(\delta q_{1},\delta p_{1},\delta X_{1},\delta Y_{1}, \delta q_{2},\delta p_{2},\delta X_{2},\delta Y_{2}\Big)^{T},
\end{equation}
the noise vector
\begin{eqnarray}\label{eq15}
\mathbf{ C}(t)=&\Big(\sqrt{2\gamma_{1}}Q_{1}^{in},\sqrt{2\gamma_{1}}P_{1}^{in},\sqrt{2\kappa_{1}}X_{1}^{in},\sqrt{2\kappa_{1}}Y_{1}^{in} \nonumber \\ &, \sqrt{2\gamma_{2}}Q_{2}^{in}, \sqrt{2\gamma_{2}}P_{2}^{in},\sqrt{2\kappa_{2}}X_{2}^{in},\sqrt{2\kappa_{2}}Y_{2}^{in}\Big)^{T},
\end{eqnarray}
and the drift matrix,
\begin{eqnarray}\label{eq14}
\mathbf{A}=
\left(
 \begin{array}{cccccccc}
-\gamma_{m1}            & \omega_{m1}   & 0            & 0               & 0              & -\eta                & 0                  & 0              \\
  -\omega_{m1} & -\gamma_{m1}      & G_{1}               & 0              & \eta            & 0                  & 0          & 0 \\
  0            & 0             & -\kappa_{1}            & \Delta_{1}     & 0              & 0                & 0                  & -\xi              \\
 G_{1}             & 0             & -\Delta_{1} & -\kappa_{1}           & 0                & 0                  &\xi          & 0 \\                                                                                     0            & -\eta             & 0            & 0               & -\gamma_{m2} & \omega_{m2}  & 0              & 0           \\
\eta       & 0             & 0            & 0               & -\omega_{m2}    & -\gamma_{m2}& G_{2}            & 0              \\
  0            & 0             & 0            & -\xi               & 0              & 0             & -\kappa_{2}  & \Delta_{2}   \\
  0            & 0             & \xi        & 0               & G_{2}             & 0                & -\Delta_{2}      & -\kappa_{2}
  \end{array}
\right),\nonumber \\
\end{eqnarray}
where $G_{j} =2g_{j} a^s_{j}$ is the effective optomechanical coupling strength of the mode $j$ to the mechanical mode. We have used the cavity mode quadratures in this case. 
In the following part, we will examine the quantum correlation between different bipartite of the proposed scheme in the domain where the multipartite system's stability is assured.

\section{Correlation matrix and stability conditions}\label{sec4}

We consider the steady state of the correlation matrix of quantum fluctuations to investigate quantum entanglement between different bipartite in the proposed model. Because the system (\ref{eq8}) is linear and the noise operators are assumed to be Gaussian with a zero-mean Gaussian state, the system can be completely represented by the resultant covariance matrix (CM) \cite{Genes2009, Genes}.
\begin{equation}\label{eq11}
\nu_{lm}=\frac{\langle \mu_{l}^{(ss)} \mu_{m}^{(ss)}+\mu_{m}^{(ss)}\mu_{l}^{(ss)}\rangle}{2},
\end{equation}
where $\mu_{l}^{(ss)}$ is the steady-state value of the $l^{th}$ component of the quadrature fluctuations vector $\boldsymbol{\mu}(t)$

The stationary CM ($\boldsymbol{\nu}$) can be obtained by solving the Lyapunov equation \cite{Vitali2007, Mari, Asjad2016, Du2017, Li2015a, Mari2012}
\begin{equation}\label{eq16}
    \mathbf{A}\boldsymbol{\nu}+\boldsymbol{\nu}\mathbf{A}^T=-\mathcal{Q},
\end{equation}
which is a linear equation for $(\boldsymbol{\nu})$ and can be directly solved, but the general rigorous expression is too cumbersome and will not be reported here. The diffusion matrix $\mathcal{Q}$ is the matrix of the steady state noise correlation functions, which is given by
\begin{eqnarray}
\label{eq17}
\mathcal{Q}=&diag \Big [ \gamma_{m1}(2\bar{n_1}+1),\gamma_{m1}(2\bar{n_1}+1),\kappa_{1},\kappa_{1} \nonumber \\ &,\gamma_{m2}(2\bar{n_2}+1), \gamma_{m2}(2\bar{n_2}+1),\kappa_{2},\kappa_{2}\Big].
\end{eqnarray}

\begin{figure}[h]
\includegraphics[width=0.32\linewidth,height=0.5\linewidth]{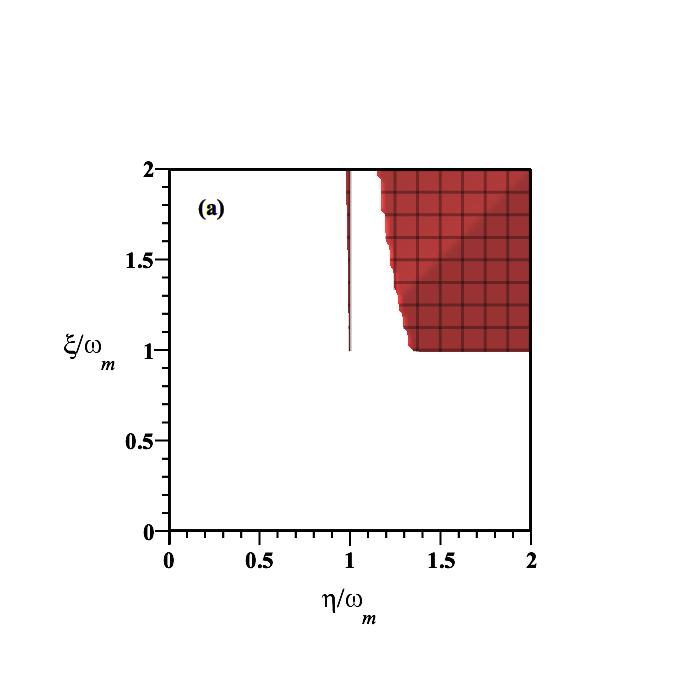}
\includegraphics[width=0.32\linewidth,height=0.5\linewidth]{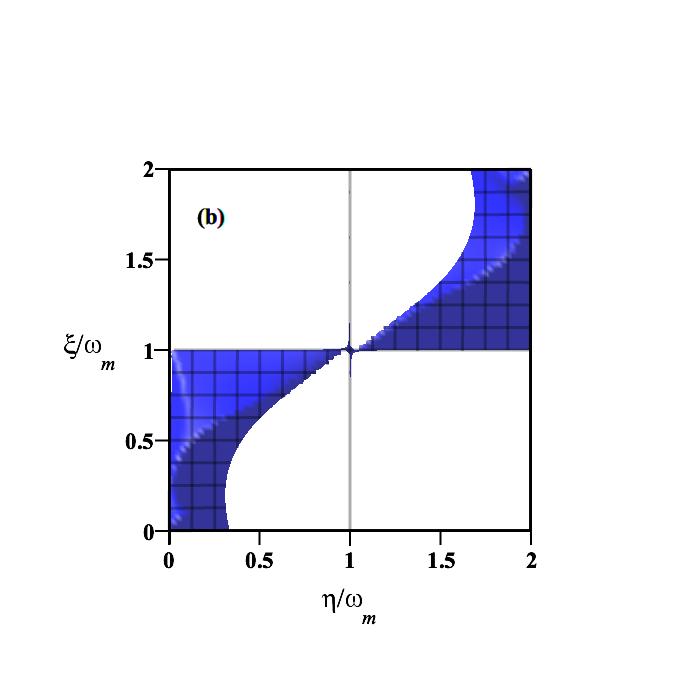}
\includegraphics[width=0.32\linewidth,height=0.5\linewidth]{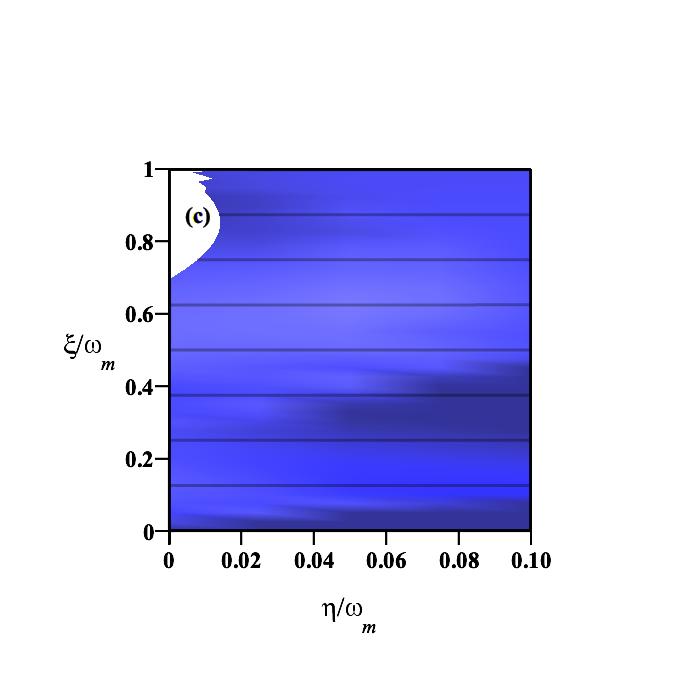}
\caption{(Color online) The maximum of the eigenvalues (real parts) of the drift matrix $A$ in terms of the normalized photon and phonon hopping couplings $\xi/\omega_m$ and $\eta/\omega_m$. (a) $\Delta \simeq -\omega_m$ red region and (b) $\Delta \simeq \omega_m,$ blue region. The color area denotes that the system is stable. We take an optimal parameters $\gamma_m/\omega_m \simeq 10^{-5}, \kappa/\omega_m \simeq 1.4$ and $g/\omega_m \simeq 2.1\times 10^{-5}$.(c) Same as in (b) for $\eta/\omega_m<0.1$ }\label{Fig2}
\end{figure}

We note that the system turns out to be stable only as all the eigenvalues of the drift matrix \textbf{A} have negative real parts. The Routh-Hurwitz criterion \cite{peres1996separability} can be used to find the parameter regime where stability emerges, although inequalities are significantly involved in the current multipartite system. As a result, we numerically plot the maximum of the eigenvalues (real parts) of the drift matrix $(\textbf{A})$ in Fig. \ref{Fig2} to keep a sensitive representation.
The maximum is bigger than zero in the white area, indicating that at least one eigenvalue has a positive real part, and so the system is unstable in this region. Under such high couplings, the "optical-spring" effect is likely to play a critical role. We consider the case of two identical optomechanical cavities for the sake of simplicity and generality, i.e.  $\kappa_1=\kappa_2=\kappa$, $G_1=G_2=G$, $\gamma_{m1}=\gamma_{m2}=\gamma_m$ and $\Delta_1=\Delta_2=\Delta$. Figures (\ref{Fig2}) illustrate the stable regime for quantum correlations under the optimal condition $\Delta’=\pm \omega_m’$ . As a consequence, this will offer a stable parameter region for experimental investigations. The color area denotes the region of the system where it is stable.  For the red detuning, it is obvious from Fig. (\ref{Fig2}-a) that $\eta/\omega_m \ge1.5$ and $\xi/\omega_m >1$.  There are two different zones of stability for the blue detuning (Fig.\ref{Fig2}-b).  In particular, (Fig.\ref{Fig2}-c) indicates that for $\eta \to 0$ the photon coupling strength may be described in the domain $0\le \xi/\omega_m\le 0.7$ for $\eta \to 0$, and  that as $\eta$ grows, the interval of $\xi$ enlarges as $0\le \xi/\omega_m\le 1$.  This finding reveals an intriguing advantage of employing phonon coupling between mechanical components: the phonon coupling enlarges the domain of the coupling strength between the cavity modes, which can extend the region for elaborating quantum concepts.

It is worth noting that the proposed scheme's appropriate potential setups could be realized with current experiments, such as the experimental system with two Fabry–Perot optical cavities, whispering cavities \cite{Guo2014}, a two-dimensional homogeneous array of optomechanical crystals \cite{Ludwig2013}, and coupled-mode heat engine in a two-membrane-in-the-middle cavity optomechanical system \cite{sheng2021realization} with a stoichiometric silicon nitride membrane (Norcada) silicon nitride membranes inside of a Fabry–Perot optical cavity.
Furthermore, we consider the parameter regime is extremely similar to the existing experimental results \cite{Ockeloen-Korppi2018a, Chille:2015us}, where $ \omega_{mj} /2\pi = 10\;\emph{MHz}, \gamma_{mj} /2\pi \simeq 100\;\emph{Hz}, \kappa_j /2\pi \simeq 2 \sim 15\; \emph{MHz}, T = 0.4 \sim 20\;\emph{K}, g_0\simeq  1.35 \emph{KHz}, \lambda_j = 810\;\emph{nm}\big). $

\section{Quantum Measures and Discussion}\label{sec5}

To quantify the quantum correlation between different bipartite of the system, we extract the corresponding reduced covariance matrix $\boldsymbol{\nu_R}$ by removing in $\boldsymbol{\nu}$ the rows and columns related to the other modes, then we get 

\begin{eqnarray}\label{eq20}
\boldsymbol{\nu_R}=
\left(
\begin{array}{cc}
\boldsymbol{\nu_1}          &\boldsymbol{\nu_c}    \\
\boldsymbol{\nu_c} ^T         & \boldsymbol{\nu_2}     \\                                                            
\end{array}\right),
\end{eqnarray}
with $\boldsymbol{\nu_1} , \boldsymbol{\nu_2} $ and $\boldsymbol{\nu_c} $ being $2\times 2$ block matrices. The two first ones represent the autocorrelations, while the last one describes the cross-correlation of the two bipartite modes. With this basis, we are now in a position to explore the quantum entanglement between different bipartite modes with various techniques.
 
It is worth exploring now that there are two main obstacles to entanglement. One is to select a method for determining if a given state in an arbitrary dimensions quantum system is separable, and the other is to determine the optimum measure for computing a given state's level of entanglement. Various metrics of entanglement and criteria for separability have been proposed to investigate these issues  \cite{bougouffa2010entanglement}.  

A good entanglement measure, on the other hand, must meet several criteria. If the states are separable, for example, the measure is zero, whereas the Bell states characterize maximally entangled states for pure states. The measure must provide unity, although different measures can produce different levels of entanglement for mixed states. There are some issues in categorizing the states based on different metrics. Furthermore, the maximally entangled mixed state is difficult to characterize. The different measures we will explore here, in particular, all have the same dimensionality, so it is a reasonable question to deal with them, especially since each one is linked to a distinct definition of entanglement.
In the subsequent subsections, we will examine some of the specific characteristics of various metrics that are significant in optomechanical research and help to fill a gap in the field of entanglement research.

\subsection{ Logarithmic Negativity}

We begin by investigating the standard metric, logarithmic negativity, which may be thought of as a quantitative version of Peres' separability criterion\cite{peres1996separability}. As a consequence, it is a full entanglement monotone under local operations and classical communication, \cite{Plenio:2005ws}, and it establishes an upper bound for the distillable entanglement,\cite{Vidal}. It has been proposed as a good measure of entanglement for Gaussian modes by \cite{Hartmann2008}  and is commonly used as an entanglement measure. The logarithmic negativity can be defined as
\begin{equation}\label{eq18a}
    E_N= \max\Big[0,-\ln(2\vartheta^{-})\Big].
\end{equation}
Here, $\vartheta^{-}$ is the smallest symplectic eigenvalue of partial transpose $\boldsymbol{\nu_R}$ matrix and is given by
\begin{equation}\label{eq19}
    \vartheta^{-}= \frac{1}{\sqrt{2}}\big(\chi(\boldsymbol{\nu_R})-[\chi(\boldsymbol{\nu_R})^2-4\det(\boldsymbol{\nu_R})]^{1/2}\big)^{1/2},
\end{equation}
and $\chi(\boldsymbol{\nu_R})\equiv \det( \boldsymbol{\nu_1})+\det (\boldsymbol{\nu_2})-2\det( \boldsymbol{\nu_c})$ ,with $\boldsymbol{\nu_1}, \boldsymbol{\nu_2}$ and $\boldsymbol{\nu_c}$ being 2$\times 2$ block matrices.

Now, using this proposed scheme within the experiment parameter domain, we investigate the impact of couplings between various modes, such as photon and phonon hopping processes, on quantum correlations and entanglement inside each cavity between the field and moving mirror.

\begin{figure}[h]
\includegraphics[width=0.8\linewidth,height=0.7\linewidth]{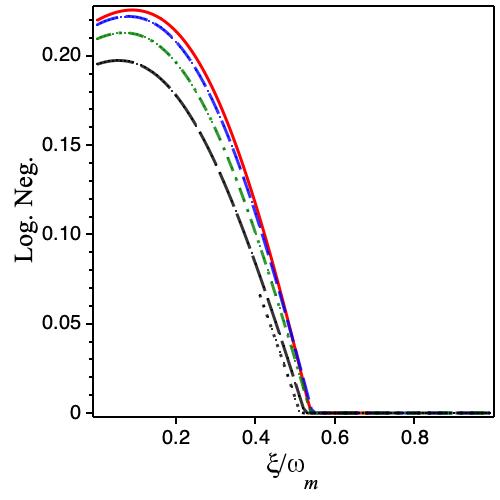}
\caption{(Color online)The steady-state logarithmic negativity  versus the normalized photon coupling strength  $\xi/\omega_m$, for different values of the  normalized phonon coupling strength  $\eta/\omega_m$.  $\eta/\omega_m=0$ (red solid line),  $\eta/\omega_m=0.1$(dashed blue line),  $\eta/\omega_m=0.2$ (dat-dashed green line),  $\eta/\omega_m=0.3$ ( long dashed black line) and  $\eta/\omega_m=0.35$ ( dotted black line).
The parameters are chosen to be ($L=1\;  \emph{mm}, m_{j} = 5\; \emph{ng} , \omega_{m}/2\pi = 10\;  \emph{MHz}, \gamma_{m}/2\pi =100\;  \emph{Hz}, \kappa/2\pi = 14\;  \emph{MHz} , \lambda =810\; \emph{nm}, \Delta \simeq \omega_m,$ $\bar{n}=836 (T=0.4K)$ and $P=35 \emph{mW}$. Where $E=\sqrt{\frac{2P\kappa}{\hbar \omega_L}}$ and $g=\frac{\omega_c}{L}\sqrt{\frac{\hbar}{m\omega_m}}$}\label{Fig3}
\end{figure}
We will assume that the two cavities are identical and that they are related via the photon and phonon hopping mechanism in the following sections. A waveguide can be used to establish photon hopping between the two optomechanical cavities, and the quantum correlation between the mechanical resonator and optical mode can be explored in the suggested scheme.

We illustrate in Fig. \ref{Fig3} the logarithmic negativity between the field and mechanical modes in one cavity in terms of the normalized photon coupling strength $\xi/\omega_m$ for different values of the normalized phonon coupling strength $\eta/\omega_m$ within the stability zone. 
The initial value of the entanglement where $\xi/\omega_m=0$ corresponds to the maximum value in the situation of separated cavities, and $\Delta \simeq \omega_m$ \cite{Bougouffa2020,Vitali2007}. With increasing $\xi/\omega_m$, entanglement increases until it reaches a maximum, then diminishes as $\xi/\omega_m$ grows larger, and finally dies abruptly. Furthermore, when $\eta/\omega_m$ increases, the amplitude of entanglement decreases.

 We will consider an interesting instance to see how the results can be interpreted. We assume that the two cavities are similar for simplicity and choose the same parameters for the two mechanical resonators and the optical modes, i.e.,  $\kappa_1=\kappa_2=\kappa$, $g_1=g_2=g_0$, and $\Delta_1=\Delta_2=\Delta$, and $\gamma_{m1}=\gamma_{m2}=\gamma_m$. Thus, the general equations (\ref{eq8}) can be written as follows:
 \begin{eqnarray}\label{eq21}
    \dot{\delta b}&=&-\beta\delta b + ig_0a^s  (\delta a^{\dag}+\delta  a)+\sqrt{2\gamma}b^{in}, \nonumber\\
    \dot{\delta a}&=&-\alpha'\delta a +i g_0a^s (\delta b^\dagger+ \delta b) +\sqrt{2\kappa} a^{in},
\end{eqnarray}
where  $\delta a=\frac{1}{\sqrt{2}}( \delta a_1+ \delta a_2)$, $\delta b=\frac{1}{\sqrt{2}}( \delta b_1+ \delta b_2)$, $\beta= \gamma_m+i(\omega_m-\eta)$,  and $\alpha'=\kappa+i(\Delta-\xi)$. Eqs. (\ref{eq21}) are the quantum Langevin equations of one optomechanical cavity with an effective detuning $(\Delta-\xi)$, where $b^{in}$ is the quantum brownian  stochastic force with zero mean value, while $a^{in}$ are the input noise operators, which satisfy the correlations relations Eqs.(\ref{eq5}, \ref{eq3}), respectively. 

\begin{figure}
\begin{center}
\includegraphics[width=0.8\linewidth,height=0.5\linewidth]{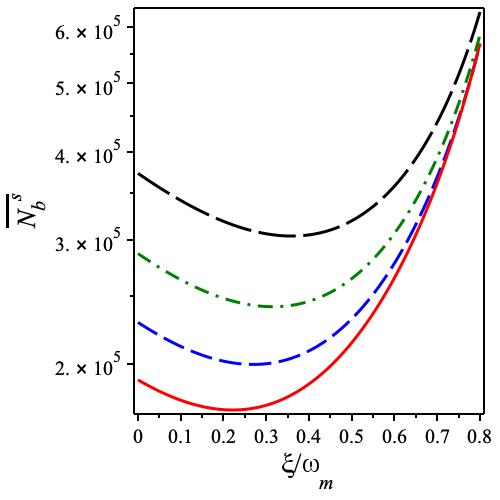}
\caption{ Final steady state average phonon number $\overline{N^s_{b}}$ in terms of the coupling strenght $\xi/\omega_m$ for the=
effective normalized detuning $\Delta/\omega_m=1$, and  for different values of the ratio $\eta/\omega_m$  ($\eta/\omega_m=0.,  0.1, 0.2, 0.3$) with red solid, blue dashed, green dash dotted,  and blackl ong dashed lines,  respectively with $\gamma_m/\omega_m=10^{-5}$, $\kappa/\omega_m=1.4$ and $\bar{n}=836$.}
\label{FigCo}
\end{center}
\end{figure}

Within this transformation, the corresponding Hamiltonian reads
\begin{equation}\label{Eq22}
\hat{H}_{lin}=(\Delta - \xi)\delta\hat{a}^\dagger \delta\hat{a} + (\omega_m-\eta)\delta \hat{b}^\dagger \delta\hat{b}
-g_0a^s(\delta\hat{a}^\dagger+ \delta\hat{a})(\delta\hat{b}^\dagger+\delta\hat{b}),
\end{equation}
where $g=g_0a_s$ illustrates the light-enhanced optomechanical coupling strength. In particular, the reference point of the cavity field is chosen such that the mean value $a_{s}$ must be real positive. Further, the rotating terms in this Hamiltonian designate the beam splitter interaction while the counter-rotating terms describe the two-mode squeezed interaction. The operators $\delta \hat{a}$ and $\delta \hat{b}$ meet equations of motion identical to the quantum master equation

\begin{equation}\label{Eq23}
\dot{\rho}=-\frac{i}{\hbar}\big[\hat{H}_{lin},\rho\big]+\frac{\kappa}{2}\mathcal{D}[\delta \hat{a}]\rho +\frac{\gamma_m}{2}(\bar{n}+1)\mathcal{D}[\delta \hat{b}]\rho+\frac{\gamma_m}{2}\bar{n}\mathcal{D}[\delta \hat{b}^\dag]\rho,
\end{equation}
where $\mathcal{D}[\hat{o}]\rho=\big[\hat{o} \rho,\hat{o}^\dag\big]+\big[\hat{o}, \rho \hat{o}^\dag\big]$ is the standard dissipator in Lindblad form, which rests invariant under the previous canonical transformation.\\
Using the master equation (\ref{Eq23}), the evolution of the mean phonon number $\bar{N}_b=\big<\delta \hat{b}^\dag \delta \hat{b}\big>= Tr(\rho \delta \hat{b}^\dag \delta \hat{b})$ can be given by a linear system of coupled ordinary differential equations relating all the independent second order moments
\begin{eqnarray}\label{Eq24}
\Lambda=&&{}\big(\bar{N}_a,\bar{N}_b, \big<\delta \hat{a}^\dag \delta\hat{b}\big>, \big<\delta \hat{a} \delta\hat{b}^\dag\big>, \big<\delta \hat{a}\delta\hat{b}\big>, \big<\delta \hat{a}^\dag \delta\hat{b}^\dag\big>,\big<\delta \hat{a}^2\big>, \big<{\delta \hat{a}^\dag}^2 \big>, \big<\delta\hat{b}^2\big>, \big<{\delta \hat{b}^\dag}^2\big> \big)^T,\nonumber\\
\end{eqnarray}
which satisfy the equation of motion that is derived in \cite{Al_Awfi_2018} and can be written as 
\begin{equation}
\label{Eq25 }
\dot{\Lambda}(t)= \mathbf{B}\Lambda(t)+\mathbf{D},
\end{equation}
where the drift matrix $B$ and the non-homogeneous  vector are given as 

\begin{eqnarray}\label{eq26}
\mathbf{B}=
\left(
 \begin{array}{cccccccccc}
-\kappa  & 0 & -i \,g  & i \,g  & -i \,g  & i \,g  & 0 & 0 & 0 & 0 
\\
 0 & -\gamma_m  & Ii\,g  & -Ii\,g  & -i \,g  & I \,g  & 0 & 0 & 0 & 0 
\\
 -I \,g  & i \,g  & \mathrm{a3} & 0 & 0 & 0 & 0 & I \,g  & -i \,g  & 0 
\\
 I \,g  & -i \,g  & 0 & \mathrm{a4} & 0 & 0 & -i \,g  & 0 & 0 & i \,g  
\\
 I \,g  & i \,g  & 0 & 0 & \mathrm{a5} & 0 & i \,g  & 0 & Ii\,g  & 0 
\\
 -Ii\,g  & -i \,g  & 0 & 0 & 0 & \mathrm{a6} & 0 & -i \,g  & 0 & -i \,g  
\\
 0 & 0 & 0 & 2\,i \,g  & 2\,i \,g  & 0 & \mathrm{a7} & 0 & 0 & 0 
\\
 0 & 0 & -2\,i \,g  & 0 & 0 & -2\,i \,g  & 0 & \mathrm{a8} & 0 & 0 
\\
 0 & 0 & 2\,i \,g  & 0 & 2\,i \,g  & 0 & 0 & 0 & \mathrm{a9} & 0 
\\
 0 & 0 & 0 & -2\,i \,g  & 0 & -2\,i \,g  & 0 & 0 & 0 & \mathrm{a10} 
  \end{array}
\right),\nonumber \\
\end{eqnarray}
and
\begin{eqnarray}\label{eq27}
\mathbf{D}=\Big(0,\gamma_{m}\bar{n},0,0, ig, -ig,0,0,0,0\Big)^{T},
\end{eqnarray}
where 
\begin{eqnarray}\label{eq28}
a3=&-\frac{\kappa+\gamma_m}{2}+i(\Delta+\omega), &a4=-\frac{\kappa+\gamma_m}{2}-i(\Delta+\omega), \nonumber \\
a5=&-\frac{\kappa+\gamma_m}{2}+i(\Delta-\omega), &a6=-\frac{\kappa+\gamma_m}{2}-i(\Delta-\omega),\nonumber\\
a7=&-\gamma_m -2i\omega, &a8=-\gamma_m+2i\omega,\nonumber\\
a9=&-\kappa+2i\Delta, &a10=-\kappa-2i\Delta,
\end{eqnarray}
As we are interested to the steady state solution $\bar{N}_{b}^s$ for the cooling mechanism, the expressions are very cumbersome, we present in Fig.~\ref{FigCo}, the variation of $\bar{N}_{b}^s$ in terms of the coupling stenghts $\xi$ and $\eta$. We consider the same parameters as in the previous figuer.
This result also indicates the parameter ranges where the final steady state average phonon number is lowest, indicating cooling in the optomechanical cavity. We have shown that the stationary intracavity entanglement in the system can be enhanced in the same region of the ground state cooling, and decreases to zero when the optomechanical cavity is heated. 

\subsection{Quantum Steering}
The measure Gaussian steering, which was recently suggested in \cite{Kogias2015} as a necessary and sufficient criterion for subjective bipartite Gaussian modes, is a fascinating alternative quantum correlation quantifier in Gaussian modes. The quantum mechanical aspect of steering is that it allows one party, $(b_1)$, to vary the state of a remote party, $(b_2)$, by altering their shared entanglement.

 Indeed, the Gaussian quantum steering in two directions in terms of the covariance matrix is described by 

\begin{eqnarray}\label{eq24}
\mathcal{G}^{b_1\to b_2} &= \max \Big(0, \mathrm{S}(2\boldsymbol{\nu_1}) - \mathrm{S}(2\boldsymbol{\nu_R})\Big), \nonumber\\
\mathcal{G}^{b_2\to b_1} &= \max \Big(0, \mathrm{S}(2\boldsymbol{\nu_2} ) - \mathrm{S}(2\boldsymbol{\nu_R})\Big), 
 \end{eqnarray}
with $\mathrm{S}(\sigma) = \frac{1}{2} \ln(\det (\sigma))$ being the R\'enyi-2 entropy \cite{Neyman1961}. 

We can distinguish the following crucial cases:
\begin{itemize}
  \item $\mathcal{G}^{b_1\to b_2}=\mathcal{G}^{b_2\to b_1}$, no steering can be observed, i.e., (no-way steering).
  \item $\mathcal{G}^{b_1\to b_2}> 0$ or $\mathcal{G}^{b_2\to b_1}>0$, one-way steering occurs.
  \item $ \mathcal{G}^{b_1\to b_2}> 0$ and $\mathcal{G}^{b_2\to b_1}>0$, shows that the bipartite modes described by the covariance matrix $\boldsymbol{\nu_R}$ are steerable from mode $b_1 (b_2)$  by Gaussian measures on mode $b_2(b_1)$, i.e., two-way steering occurs.
\end{itemize}
Remarkably, the expression $\mathcal{I}^{b_1<b_2}=\mathrm{S}(2\boldsymbol{\nu_1}) - \mathrm{S}(2\boldsymbol{\nu_R})$ can be perceived as a form of quantum coherent information \cite{wilde2013quantum}, but with Rényi-2 entropies substituting the predictable von Neumann entropies. 

Furthermore, the Gaussian steering measure satisfies the following valuable properties for mode Gaussian states: (a) $\mathcal{G}^{b_1\to b_2}$ is convex, (b) $\mathcal{G}^{b_1\to b_2}$ is monotonically decreasing under quantum operations on the untrusted steering party $(b_1)$; (c) $\mathcal{G}^{b_1\to b_2}$ is additive; (d) $\mathcal{G}^{b_1\to b_2} =\mathcal{E}(\nu_R)$ for $\nu_R$ pure, and (e) $\mathcal{G}^{b_1\to b_2} \leq \mathcal{E}(\nu_R)$ for $\nu_R$ mixed, where $\mathcal{E}(\nu_R)$ means the Gaussian Rényi-2 measure of entanglement \cite{adesso2012measuring}.
The features of stationary quantum steering are then investigated, as well as how to achieve one-way quantum steering by altering coupling strengths and identifying the appropriate region of stability.  The Gaussian steerability takes on a primarily simple form when each party has only one mode and the covariance matrix Eq. \ref{eq20}. The Shur complement of the mode $a_1$ in the covariance matrix $\nu_R$, given as $M_{\nu_R}^{M}=\nu_2-\nu^T\nu_1^{-1}\nu_c$,  has a single symplectic eigenvalue,  $\bar{\nu}_{M}=\sqrt{det(M_{\nu_R}^{M})}$. . After that, the cavity mode's Gaussian measurement is given as
\begin{equation}
\label{eq30}
\mathcal{G}^{F\to M} = \max \Big(0, - \ln(\bar{\nu}_{M})\Big).
\end{equation}

By swapping the roles of $F$ and $M$, an equivalent measure of Gaussian  $M → F$ steerability can be obtained, resulting in an equation like Eq.\ref{eq30}, in which the symplectic eigenvalues of the $2\times 2$ Schur complement of $F$, $M_{\nu_R}^{F}=\nu_1-\nu\nu_2^{-1}\nu_c^T$,appear instead.

\begin{figure}[ht]
\includegraphics[width=0.5\linewidth,height=0.5\linewidth]{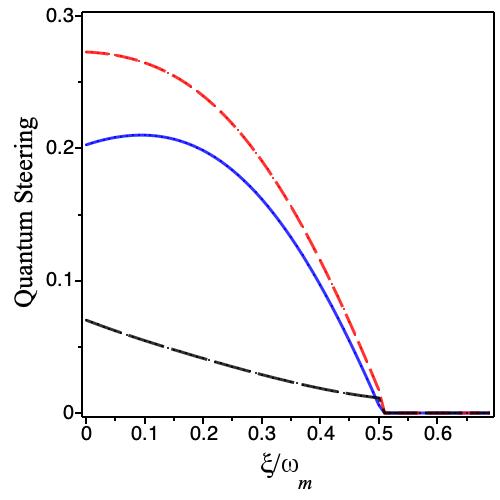}
\includegraphics[width=0.5\linewidth,height=0.5\linewidth]{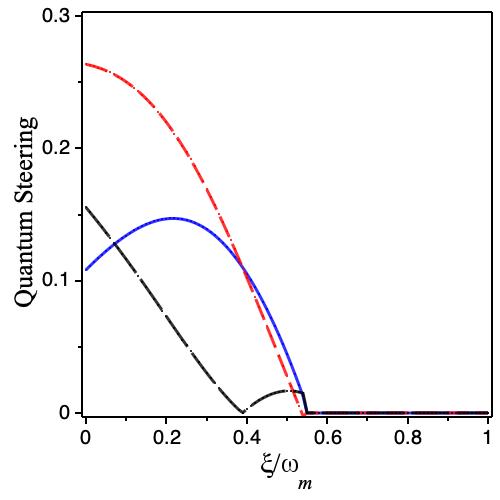}
\caption{(Color online)(a)The steady-state quantum steering  $\mathcal{G}^{F\to M}$ (blue solid line), $\mathcal{G}^{M\to F}$ (red dashed line), and the asymmetry of steering $|\mathcal{G}^{M\to F}-\mathcal{G}^{F\to M}|$(black long dashed line) in terms of the normalized photon coupling strength  $\xi/\omega_m$, for different values of the  normalized phonon coupling strength  $\eta/\omega_m$.  $\eta/\omega_m=0$ (left panel),  $\eta/\omega_m=0.2$(right panel). The other parameters are the same as in ﬁgure \ref{Fig3}. }\label{Fig5}
\end{figure}

For different values of the normalized phonon coupling strength $\eta/\omega_m$, we show the change of quantum steering against the normalized photon coupling strength $\xi/\omega_m$ in Figure \ref{Fig5},. For normalized photon coupling strengths less than a threshold value $xi 0/omega m$, both Gaussian steering $\mathcal{G}^{F\to M}$ and $\mathcal{G}^{M\to F}$  between the field and mechanical modes in one cavity are positive. The bipartite modes can then be steered from mode $Field(Mirror)$ using Gaussian measures on mode $Field(Mirror)$, resulting in two-way steering. While both quantum steering $\mathcal{G}^{F\to M}$ and $\mathcal{G}^{M\to F}$ become zero in the same region where entanglement is also zero for normalized photon coupling strengths greater than a threshold value  $\xi_0/\omega_m$,  this means that steering requires entanglement, i.e. no steering without entanglement.

There can be no one-way steering within the chosen parameter ranges, therefore we have $\mathcal{G}^{F\to M}\ne \mathcal{G}^{M\to F}$. Except for one $\xi$ when $\eta \ne 0$. Furthermore, as the normalized phonon coupling strength  $\eta/\omega_m$ increases, the threshold value of the normalized photon coupling strength increases, implying that entanglement between the cavity and mechanical modes is always present. We can check the steering asymmetry of two-mode Gaussian states, which can be defined as  $|\mathcal{G}^{F\to M} - \mathcal{G}^{M\to F}|$, because $\mathcal{G}^{F\to M} \ne \mathcal{G}^{M\to F}$. In \cite{handchen2012observation}, the asymmetry of steering in the Gaussian context was experimentally proven.

As a result, steering and entanglement are nonmonotonic functions of ($\xi$ and $\eta$), with extreme values for specific coupling strengths. The obtained results show that as the phonon coupling strength increases, the ideal values of both entanglement and steering decrease. Furthermore, when the photon coupling strength is high enough, entanglement and steering evaporate completely, which corresponds to system heating, as previously explained. For relatively large values of coupling strengths, where the logarithmic negativity is always distinct from zero, the steering can arise. This shows that entanglement is a nonclassical correlation that is stronger than steering.

\section{Conclusions}\label{sec6}
Finally, we have systematized a hybrid optomechanics system consisting of two optomechanical cavities in which the optical modes are connected via photon hopping and the mechanical resonators via phonon tunneling. 
The output classical field is directed at both cavities.  We have provided two different quantum measures that can be utilized with this scheme and analyzed their implications.
Furthermore, we have demonstrated that the intracavity entanglement is extremely sensitive to the coupling strengths established between the connected cavities on both sides. In addition, the stability conditions are thoroughly investigated using the most recent experimentally available values. The degree of intracavity entanglement is tested using two distinct techniques, namely logarithmic negativity and quantum steering, again when the stability criteria are achieved. A convenient choice of coupling strengths can improve intracavity entanglement, as we have illustrated. As a result, establishing phonon and photon hopping processes between cavity optomechanics can improve quantum correlations in the examined domain, which is the unresolved sideband Doppler regime $\kappa > \omega_m$. We have established that the generated entanglement appears in the same region of the ground state cooling. 
In the future, other regimes \cite{sun2022conditions} can be investigated in greater depth. 
The numerical computational results reveal that a convenient choice of coupling strengths established between the two optomechanical cavities can considerably influence both entanglement and steering of the mechanical and optical modes. When coupling strengths are raised, it is discovered that the $F \to M$ steering disappears before entanglement. 
On the other hand, the current inquiry is not limited to the specified modest scheme. Different quantum regimes can be investigated for the realization of quantum memory for continuous-variable quantum information processing, quantum-limited displacement measurements, and other complex types of coupled-cavity optomechanics.

\begin{acknowledgements}
This research was supported by the Deanship of Scientific Research, Imam Mohammad Ibn Saud Islamic University,
Saudi Arabia, Grant No. (20-13-12-004)\\
We thank Z. Ficek for the valuable discussions.
\end{acknowledgements} 

\bibliographystyle{apsrev4-1}
\bibliography{Ref1}

\end{document}